# Phosphorous dimerization in GaP at high pressure

Barbara Lavina[1,2,*], Eunja Kim[2], Hyunchae Cynn[3], Philippe F. Weck[4], Kelly Seaborg[1,2], Emily Siska[1], Yue Meng[5], William Evans[3]

[1]*High Pressure Science and Engineering Center, University of Nevada, Las Vegas*

[2]*Department of Physics and Astronomy, University of Nevada, Las Vegas*

[3]*Lawrence Livermore National Laboratory*

[4]*Sandia National Laboratories*

[5] *HPCAT, Carnegie Institution of Washington*

*e-mail: lavina@physics.unlv.edu

**Using combined experimental and computational approaches we show that at 43 GPa and 1300 K gallium phosphide adopts the "super-*Cmcm*" structure, here indicated with its Pearson notation *oS*24. First-principles enthalpy calculations demonstrate that this *oS*24 structure is more thermodynamically stable above ~ 20 GPa than other proposed polymorphs, such as cubic ZB and SC16 or the ubiquitous orthorhombic *Cmcm* phase (*oS*8). It is found that *oS*24-structured GaP exhibits short P-P bonds, according to our high-resolution structural analysis. Such phosphorous dimerization in GaP, observed for the first time and confirmed by *ab initio* calculations, sheds light on the nature of the super-*Cmcm* structure and provides critical new insights into the high-pressure polymorphism of octet semiconductors. Bond directionality and anisotropy, in addition to 5-fold phosphorous coordination, explain the relatively low symmetry of this high-pressure structure.**

**Introduction**

The high-pressure behavior of semiconductors has been extensively investigated since the discovery of pressure-induced sharp resistivity decreases across this class of compounds, indicative of transitions to metallic states[1,2]. Studies of the associated structural changes followed, with the aim of linking atomic arrangements to physical properties and uncovering the systematic behavior of this important class of materials. It was originally thought that increasing pressures would induce an evolution of the bonding character from partially covalent to purely ionic, and eventually metallic[3,4]. The corresponding sequence of phase transitions was expected to be zinc blend →rock salt → diatomic $β$-Sn. This paradigm was challenged as the resolution of powder diffraction analysis significantly improved and several low symmetry structures were revealed. An overall consensus on the polymorphism of group IV, III-V and II-VI semiconductors was reached more than a decade ago[3–5]. Our findings suggest that there is more to learn on the nature of semiconductors polymorphism.

Gallium phosphide (GaP) crystallizes in the zinc blend structure (ZB) and features a pressure-induced transition to a metallic state at ~22 GPa[2]. It was shown that metallic GaP assumes an orthorhombic structure, with Pearson symbol $oS8$, which can be described as a distorted NaCl arrangement[6], dismissing the previously proposed $β$-Sn structure[7]. While $β$-Sn type high-pressure GaP phases have been a subject of considerable debate[8,9], a small stability range for the SC16 arrangement (simple cubic SC16) between the ambient pressure ZB phase and the *Cmcm* polymorph was suggested for GaP[10–12]. Yet much attention has been focused on the cubic ZB → *Cmcm* structural transition occurring around 22 GPa[4–6,10]. Nelmes *et al.*[6] proposed that both GaP and GaAs subjected to high pressure crystallize in the site-disordered $oS8$ structure, while AlSb, InP, InAs, CdTe, HgTe, and HgSe adopt its site-ordered counterpart. Diffraction analysis suggests that $oS8$ GaP is a long-range, site-disordered structure[6,13], whereas extended X-ray absorption fine structure (EXAFS) data recorded at the gallium $K$-edge ($E$ = 10.367 keV) show that *Cmcm* GaP is a short-range ordered structure with six Ga-P bonds and no Ga-Ga bonds at 39 GPa[13]. A recent study suggests a nearly isobaric boundary at ~ 22 GPa between ZB and $oS8$ phases up to ~ 800 K[14]. In spite of the wealth of experimental and theoretical investigations of the ZB → *Cmcm* transition, it remains unclear if site-disordered arrangements exist in the *Cmcm* GaP phase, which is of crucial importance to understand the nature of phase transitions in this class of octet semiconductors.

In this study, the high-pressure behavior of GaP was investigated under quasi-hydrostatic conditions using single-crystal and multigrain X-ray diffraction (XRD) techniques, combined with density functional theory (DFT) calculations. The present high-resolution crystallographic analysis, corroborated by DFT modeling, allows us to explain the lattice anisotropy of the GaP high-pressure structure. Specifically, we show that GaP adopts the "super-*Cmcm*" structure, with Pearson notation $oS24$, at $P$ = 43 GPa and $T = 1300$ K, which exhibits short P-P bonds indicative of dimerization. The relationship between the two *Cmcm* GaP phases (*oS*8 and *oS*24) is also revisited.

**Experimental and theoretical methods**

The specimen used in this study was a thin, light-brown colored, single-crystal wafer. Trace metals analysis with inductively coupled plasma mass spectrometry (ICP-MS) determined the sample contained 34.5 µg/g of boron and no other detectable contaminants. Samples were compressed using a diamond anvil cell (DAC) and probed via XRD at the high-pressure beam line 16-IDB of HP-CAT, Advanced Photon Source, Argonne National Laboratory. Between the anvils' culets, 300 µm in diameter, a rhenium gasket indented to ~ 40 µm thickness with a 120 µm hole provided the sample chamber. A fragment of the specimen along with gold fine powder and pre-pressurized neon gas[15] were loaded in the sample chamber. Pressure was determined using Au EOS[16]. The sample was annealed at 43 GPa using an online double-sided infrared (IR) laser-heating setup in which temperature is determined from the emitted radiation[17]. The sample was analyzed using a 33.169 keV X-ray beam. Diffracted beams were collected with a MAR165-CCD area detector, which was calibrated using powder patterns of $CeO_2$ standard. Single-crystal and multigrain diffraction data were collected with the rotation method, with a resolution of 0.6 Å. Data reduction was performed with the software FIT2D[18], GSE_ADA and RSV[19]. Structural solution and refinements were carried out using the Endeavour, and Shelxl software[20].

First-principles total-energy calculations were performed using spin-polarized DFT, as implemented in the Vienna *ab initio* simulation package (VASP)[21]. The exchange-correlation energy was calculated using the generalized gradient approximation (GGA) with the parameterization of Perdew, Burke and Ernzerhof (PBE)[22]. The interaction between valence electrons and ionic cores was described by the projector augmented wave (PAW) method[23,24]. The Ga ($3d^{10}$, $4s^2$, $4p^1$) and P ($3s^2$, $3p^3$) electrons were treated explicitly as valence electrons in the Kohn-Sham (KS) equations and the remaining core electrons together with the nuclei were represented by PAW pseudopotentials. The plane-wave cutoff energy for the electronic wavefunctions was set to a value of 500 eV, ensuring the total energy of the system to be converged to within $10^{-5}$ eV/atom. Periodic unit cells containing 4, 8, 4, and 12 formula units were used in the calculations for the ZB, SC16, *oS*8, and *oS*24 structures, respectively, in order to investigate the cubic ZB → orthorhombic *Cmcm* transition. Ionic relaxation was carried out using the quasi-Newton method and the Hellmann-Feynman forces acting on atoms were calculated with a convergence tolerance set to 0.001 eV/Å. The Brillouin zone was sampled using the Monkhorst-Pack special *k*-point scheme[25] with a 7x7x7 mesh for structural optimization and total-energy calculations.

**Results and Discussions**

The diffraction patterns of the ZB phase show sharp peaks typical of unstrained single-crystals up to 24.5 GPa (Fig. S1-2). The unit cell parameter was refined against the *d*-spacing of an average of ~160 peaks. The bulk modulus of the ZB phase was determined using 14 data points (Fig. 1) applying the 3$^{rd}$ order Birch-Murnaghan EOS. By fixing the first derivative to the literature value of 4.5[10], an experimental bulk modulus of 87.5 (6) GPa was obtained, in good agreement with our theoretical value of 83.5 GPa from DFT calculations, and the values of 86.4 from Soni et al.[26] and 90 GPa from Mujica and Needs[10]; these values appear slightly larger than the theoretical estimate of 78.35 GPa recently reported[12].

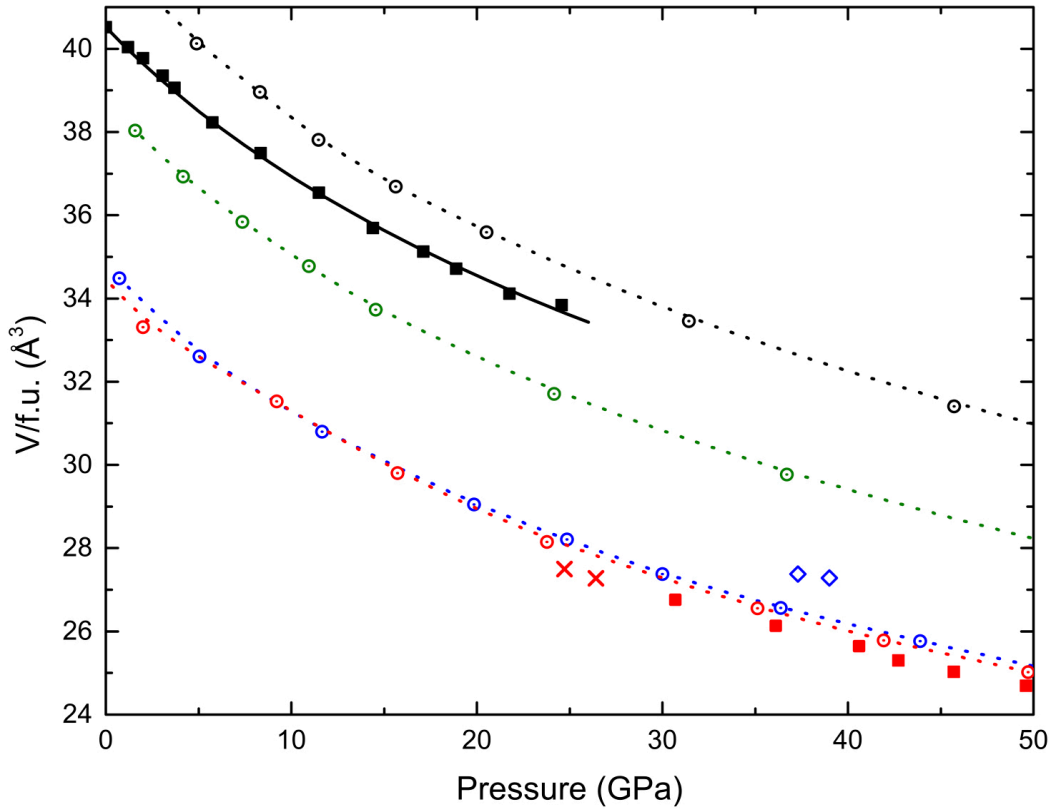

**Figure 1**. Experimental and calculated *P-V* data of GaP polymorphs. Solid black squares: ZB; black line: ZB experimental EOS; blue diamonds: *oS*8 literature[6,13]; red solid squares: *oS*24; red crosses: *oS*24 showing weak diffuse scattering lines; black, green, blue and red dotted lines: calculated volumes per formula unit of ZB, SC16, *oS*8 and *oS*24 respectively. Error bars are smaller than symbol sizes.

Above 25 GPa, diffraction patterns revealed the onset of a sluggish structural phase transition, as evidenced by severe peaks broadening and emergence of new peaks. The onset of phase transition occurs at higher pressure, by roughly 3 GPa, than most recent results[14]. Differences in stress conditions explain such apparent disagreement. Diffraction peaks of the high-pressure phase were very weak and broad, hindering the structural analysis. Hence, the sample was heated for a few minutes up to 1300 K at 43 GPa in order to improve its crystallinity. New sharp peaks appeared in the first collected diffraction pattern, about a minute into the heating, indicating that extensive recrystallization readily occurred. After quenching, we collected multigrain diffraction patterns containing about a thousand peaks from few grains of a single orthorhombic phase (Fig. S3). Grains had random mutual orientations. The *a* and *b* lattice parameters are similar to those of the *oS*8 phase, while the *c* parameter is roughly 3 times longer than the corresponding length in *oS*8 (**Table 1**). Inspection of structure factors indicates *Cmcm* as the most plausible space group. Structural solution and refinements were performed using the integrated intensities (after scaling and merging) of the largest crystals identified in two multigrain patterns collected from different sample locations[27]. We found that at ~ 43 GPa and 1300 K the sample crystallizes with the *oS*24 structure type first observed in InSb at ~ 5 GPa[28].

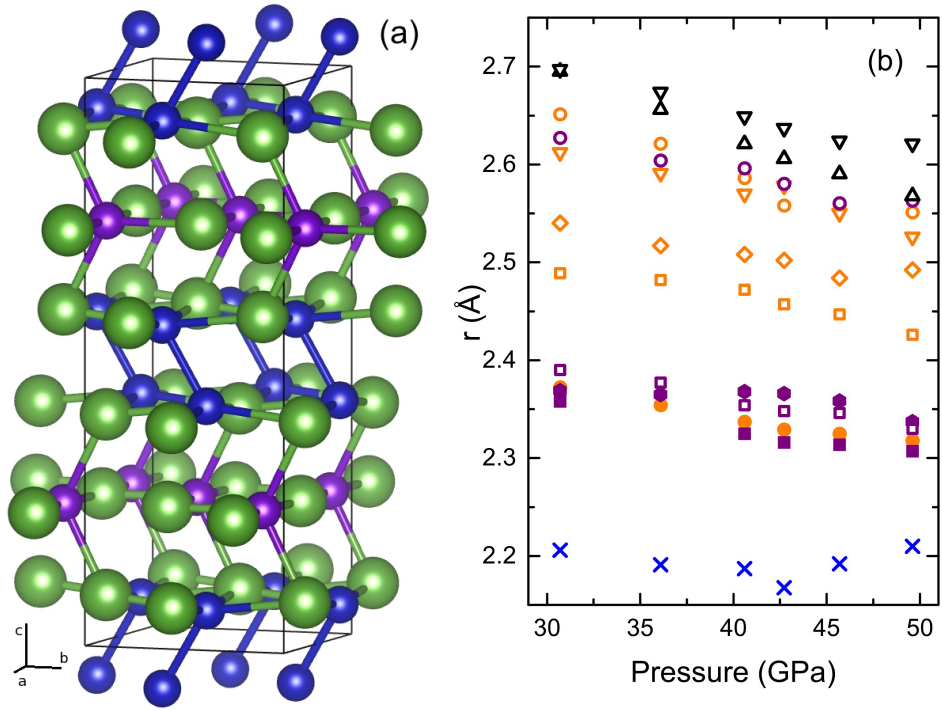

**Figure 2**. a) The structure of *oS*24 is represented with blue P1 atoms forming P-P bonds between layers and purple P2 atoms bonded to the green Ga atoms[33]. b) Interatomic distances in *oS*24 GaP. Blue crosses: P1-P1; orange symbols: P1-Ga; purple symbols: P2-Ga; black triangles: Ga-Ga. Solid symbols represent distances with multiplicity 2.

In *oS*24 phosphorous atoms shows strong bonding differentiation, P1 atoms located in the 8f site form P-P dimers (Fig. 2a) and are twice as many as the P2 atoms located in 4c crystallographic sites which are exclusively bonded to Ga atoms in a distorted 5-fold coordination. Stacked perpendicularly to the *c*-axis, double layers of distorted P1-Ga squares connected via out-of-plane P1-P1 bonds alternates with the NaCl-like single layers with the P2 atoms (point symmetry *4c*). Gallium is also arranged in two non-equivalent crystallographic sites (point symmetry *8f* and *4c*, **Table 1**). All sites are fully occupied within uncertainty (~5%), showing that the phase is largely stoichiometric and ordered. Furthermore, all observed peaks are accounted for, ruling out the occurrence of significant chemical reactions during heating. The P1-P1 interatomic distance, around 2.2 Å, compares well with values observed in several compounds[29]; and is markedly the shortest in the structure (Fig. 2b). P1 also has two Ga atoms at short distance; two at intermediate and two at long distances, resulting in highly distorted coordination geometry. P2 does not form P-P bonds, it shows five short P-Ga bonds, while a sixth Ga atom lies at greater distance (Fig. 2b). Ga-P first coordination distances can be divided in three groups (Fig.2b): shorter lengths between 2.3 and 2.4 Å (close to the ambient conditions bond length, 2.358 Å), intermediate values around 2.5 Å, and distances that might reflect weak interactions, around 2.6 Å. It follows that phosphorous shows a tendency for 5-fold coordination at these pressures in GaP. The Ga-Ga shortest distances measure about 2.6 Å, which is comparable with first coordination Ga-Ga distances found in elemental gallium at ambient conditions.

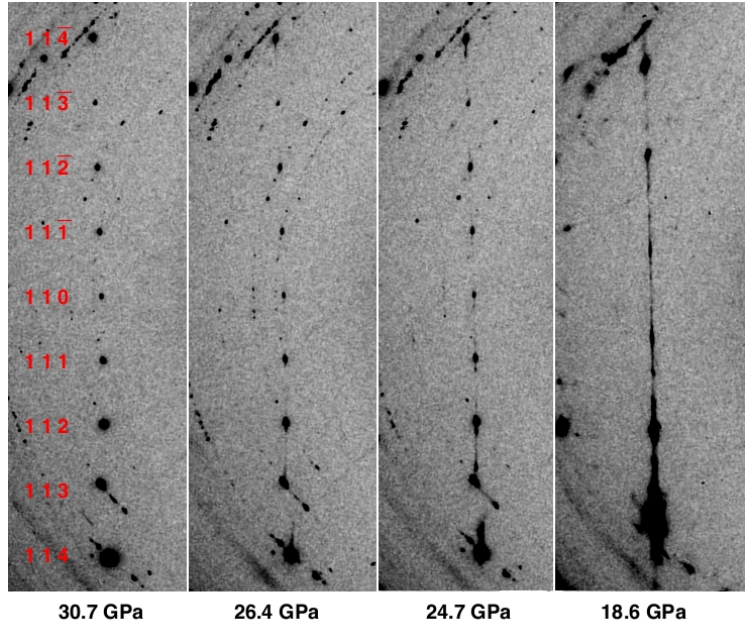

**Figure 3**. Details of diffraction patterns of *oS*24 GaP. Diffuse scattering lines between Bragg's peaks are developed in decompression.

After heating, *oS*24 GaP was compressed at ambient temperature up to ~ 50 GPa and then decompressed to 18.6 GPa. Diffraction patterns and structural refinements (Table S1-2) do not suggest further phase changes up to the highest pressure and down to ~ 30 GPa. Structural refinements indicate a predictable anisotropy of bond compressibility (Fig. 2b), featuring a rigid dimer. Diffuse scattering lines connecting rows of Bragg's peaks along the *c*-axis direction (Fig. 3) appeared in decompression starting at 26.4 GPa. These effects were subtle when first detected, but quickly gained intensity, thus demonstrating a rapid loss of long range ordering along the *c*-axis direction. It appears that, while the NaCl-like layers are preserved, they experience a relative displacement, which might be driven by a breakdown of the dimers. Hence the onset of instability of *oS*24 in decompression can be placed between 30 and 26.4 GPa. The volume difference of ZB and *oS*24 is ~18% at 25 GPa, which is similar to the value from the previous studies reporting the ZB → site-disordered *Cmcm* transition that resulted in the volume collapse of 18.4% at 26 GPa[6].

**Table 1.** Structural parameters of *oS*24 GaP from x-ray diffraction analysis and *ab initio* calculations.

|  | Experimental | Calculated |
| --- | --- | --- |
| P (GPa) | 42.7 | 42 |
| T (K) | 298 | 0 |
| *a* (Å) | 4.621 (2) | 4.679 |
| *b* (Å) | 4.927 (2) | 4.981 |
| *c* (Å) | 13.335(7) | 13.275 |
| V (Å$^3$) | 303.27 | 309.39 |
| P1 (8*f*) | 0, 0.3890(8), 0.0703(6) | 0, 0.38982, 0.07177 |
| P2 (4*c*) | 0, 0.1084(12), 1/4 | 0, 0.11056, 1/4 |
| Ga1 (8*f*) | 0, 0.1064(4), 0.5909(2) | 0, 0.10788, 0.59030 |
| Ga2 (4*c*) | 0, 0.5849(5), 1/4 | 0, 0.58698, 1/4 |

The experimental unit cell parameters were determined from least squares refinement against the *d*-spacing of 345 peaks. The refinement of atomic coordinates was performed against squared structure factors ($N_{all}$ =115, $R_{eq}$ =12%, $R_\sigma$ =7.5%). The refinement converged with satisfactory statistical parameters ($R_1$=6.5%, $R_{4\sigma}$ =6.7%, $wR_2$=17.5%).

In order to study the ground state stability of GaP at high pressure the energetics of the new polymorph was compared to those of the candidate phases in the pressure range studied, i.e. ZB, SC16 and $oS8$. Total energy and enthalpy were calculated using DFT. Good agreement between predicted and observed compressibility was obtained for all phases (Fig. 1, S4), with excellent agreement achieved also for atomic parameters (Table 1). Fig. 4(a) shows the calculated binding energy curves, suggesting several possible pressure-induced structural transitions. Calculations predict that the low-pressure ZB phase undergoes a phase transition to a high-pressure phase adopting the SC16 structure, followed by the orthorhombic phase $oS24$ at greater pressure. The narrow stability range predicted for SC16 is consistent with results from previous theoretical studies[10] and is similar to the behavior theoretically predicted and experimentally confirmed for GaAs[11]. While the ZB → SC16 transition may not be ruled out according to the aforementioned energetics calculations, we focus in this study on the cubic ZB → orthorhomic $Cmcm$ transition, with proposed $oS8$ and $oS24$ $Cmcm$ phases.

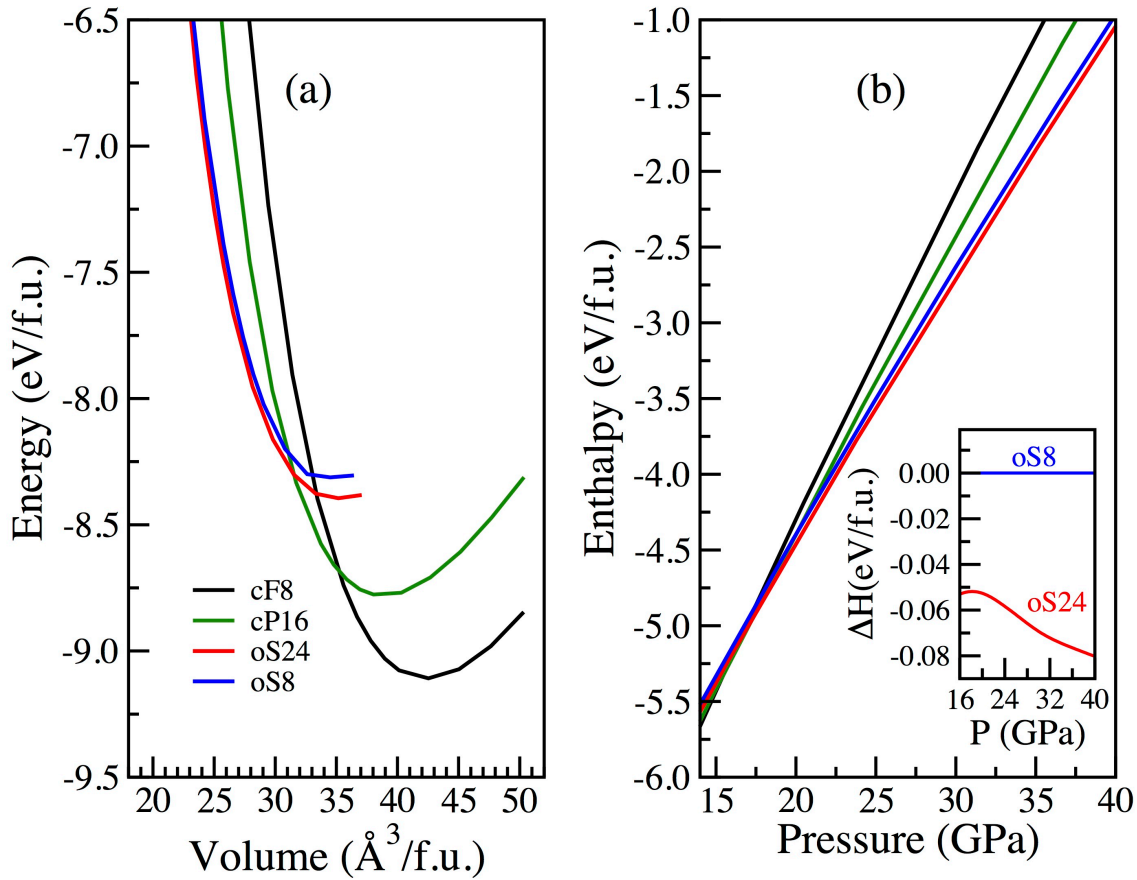

**Figure 4**. The calculated energetics of GaP: (a) total energy vs. volume and (b) enthalpy vs. volume; the inset depicts the $oS8$-$oS24$ enthalpy difference in the range most relevant to this work. The enthalpy of $oS8$ is taken as reference.

The effect of site disorder on the relative stability of the GaP polymorphs has been examined using DFT. It is found that site disordered states in $oS8$ have higher energy than the ordered state by 0.79 and 0.68 eV when a site-disorder was created in the unit cell containing 8 atoms, by exchanging the Ga and P sites randomly, at 20 and 29.3 GPa, respectively. As shown in Fig. 4(a), while $oS24$ has the lowest total energy above 18 GPa, $oS8$ and $oS24$ are almost energetically degenerate. Conversely, the comparison of enthalpies [Fig. 4(b)] clearly shows that the site-ordered $oS8$ is never energetically favored in the pressure range explored in our analysis when compared to $oS24$. Hence, our energetics analysis indicates that $oS24$ is a stable ground state polymorph of GaP and not just a high P-T phase. Furthermore, we calculated the physical properties of GaP (Table 2); we found that the band gap is closed in the two orthorhombic phases, hence they both would be consistent with resistivity measurements[1,2].

**Table 2**. The calculated physical properties of GaP. The number in parenthesis is from reference [32].

| Pressure (GPa) | 0 | | 24 | |
| --- | --- | --- | --- | --- |
| Crystal structure | ZB | SC16 | $oS8$ | $oS24$ |
| Lattice constants (Å) | 5.525 | 6.33 | 4.782 | 4.805 |
|  |  |  | 4.975 | 5.092 |
|  |  |  | 4.603 | 13.803 |
| $E_g$ (eV) | 1.65 (2.26) | 0.24 | 0.0 | 0.0 |

The onset of $oS24$ GaP instability upon decompression occurs within its calculated stability field. Furthermore, the SC16 phase was not observed experimentally, even upon heating[14]. Inconsistencies among different experiments and with calculation are very common in high-pressure semiconductors studies[3–5]. While some differences between predictions and experiments ought to be expected, the combination of rich phase diagrams with narrow stability ranges for phases such as SC16, experimental limitations (chiefly hydrostaticity and resolution, here dramatically improved), and most importantly the kinetic hindrance of the transition between all these phases, can explain seemingly inconsistent observations. The polymorph obtained upon compression without heating[6] and even with moderate heating[14] shows broad diffraction peaks and discrepancy between long and short range ordering [13]. While we do not exclude that such phase is indeed $oS8$ (metastable), it is also possible that it is a complex material with a very high defect concentration in which $oS24$-like clusters might develop, but where the extensive ordering required forming a well-crystallized $oS24$ is not achieved, even upon moderate heating. More importantly, the structural features here described for $oS24$ GaP are in sheer contrast with the "quasi-monoatomic" description often suggested for disordered high-pressure phases[6].

While shortest bonds between like-atoms were observed in $oS24$ InSb[28], the possible occurrence of dimerization was not recognized in previous studies, and such short bonds were considered as "wrong bonds", thus associated with instability[5]. It should be noted that, compared to P-P bonds in GaP, Sb-Sb bonds in InSb are not as short with respect to In-Sb bonds. In addition, the powder diffraction structural analysis that quite remarkably resulted in the discovery of the correct $oS24$ InSb structure, was not of high resolution, two fractional coordinates were indeed fixed[28]. To our knowledge, only one recent theoretical study suggested the formation of phosphorous dimers in an octet high-pressure phase of boron phosphide, although at much higher pressure[30]. Kelsey and Ackland[31] compared the $oS8$ and $oS24$ structure types and concluded that the two orthorhombic polymorphs are almost energetically equivalent, as they provide similar local configurations (i.e. number of like- and unlike-neighbor atoms). In GaP, we show that the $oS8$ and $oS24$ structures differ markedly. While the site-disordered $oS8$ most likely contains both Ga-Ga and P-P bonds, the $oS24$ GaP has no Ga-Ga bonds. The $oS24$ phase is more energetically favorable than $oS8$, for example, by 0.05 eV at 20 GPa and 0.08 eV at 40 GPa. The $oS24$ structure allows for the development of a strong bonding anisotropy. $oS24$ GaP shows a spread of Ga-P bond lengths and distorted coordination geometries, but no Ga-Ga bonds, i.e. consistent with previous EXAFS findings[13]. It can be inferred that the short P-P dimers present in $oS24$ GaP facilitate elongation of in the unit cell along the $c$-axis - in stark contrast with the $oS8$ structure that doesn't feature P-P dimers - resulting in the "super *Cmcm*" structure. The GaP case study offers an explanation for the structural complexity that might apply to other semiconductors high-pressure polymorphs.

**Conclusion**

In conclusion, we present a high-resolution structural refinement and energetic analysis of the high-pressure polymorph of the GaP semiconductor that adopts the $oS24$ structure, owing to the formation of P-P dimers and highly anisotropic bonding, which explain for the first time the relatively low symmetry of this polymorph. In the literature several semiconductors high-pressure phases were suggested to be site-disordered, and considered quasi-monoatomic. Gallium phosphide was among them. We found that the equilibrium phase is all but disordered: in $oS24$ GaP non-equivalent phosphorous atoms show strong differentiation. Our results demonstrate that simple concepts often used to rationalize the systematics of semiconductors polymorphism, such as the degree of ionicity, ionic sizes, electronegativity and the instability of like atoms bonds, cannot always account for the nature of the polymorphism of binary semiconductors.

**Acknowledgements**

This research was sponsored by the National Nuclear Security Administration under the Stewardship Science Academic Alliances program through DOE Cooperative Agreement No. DE-NA0001982. This work performed under the auspices of the U.S. Department of Energy by Lawrence Livermore National Laboratory under Contract DE-AC52-07NA27344. EK acknowledges funding supported by the National Aeronautics and Space Administration under Agreement No. NNX10AN23H issued through the NASA Space Grant. Sandia National Laboratories is a multi-mission laboratory managed and operated by National Technology and Engineering Solutions of Sandia, LLC, a wholly owned subsidiary of Honeywell International, Inc., for the U.S. Department of Energy's National Nuclear Security Administration under Contract DE-NA0003525. This work was conducted at HPCAT (Sector 16), Advanced Photon Source (APS), Argonne National Laboratory. HPCAT operations are supported by DOE- NNSA under Award No. DE-NA0001974 and DOE-BES under Award No. DE-FG02-99ER45775, with partial instrumentation funding by NSF. APS is supported by DOE-BES, under Contract No. DE-AC02-06CH11357. Use of the COMPRES-GSECARS gas loading system was supported by COMPRES under NSF Cooperative Agreement EAR 11-57758 and by GSECARS through NSF grant EAR-1128799 and DOE grant DE-FG02-94ER14466. We thank Sergey Tkachev for his assistance with the gas loadings. The sample was kindly made available for this study by Prof. Malcolm Nicol.

# Supplementary Information: Phosphorous dimerization in GaP high-pressure polymorph

Barbara Lavina[1,2], Eunja Kim[2], Hyunchae Cynn[3], Philippe F. Weck[4], Kelly Seaborg[1,2], Emily Siska[1], Yue Meng[5], William Evans[3]

[1]*High Pressure Science and Engineering Center, University of Nevada, Las Vegas*
[2]*Department of Physics and Astronomy, University of Nevada, Las Vegas*
[3]*Lawrence Livermore National Laboratory*
[4]*Sandia National Laboratory*
[5] *HPCAT, Carnegie Institution of Washington*

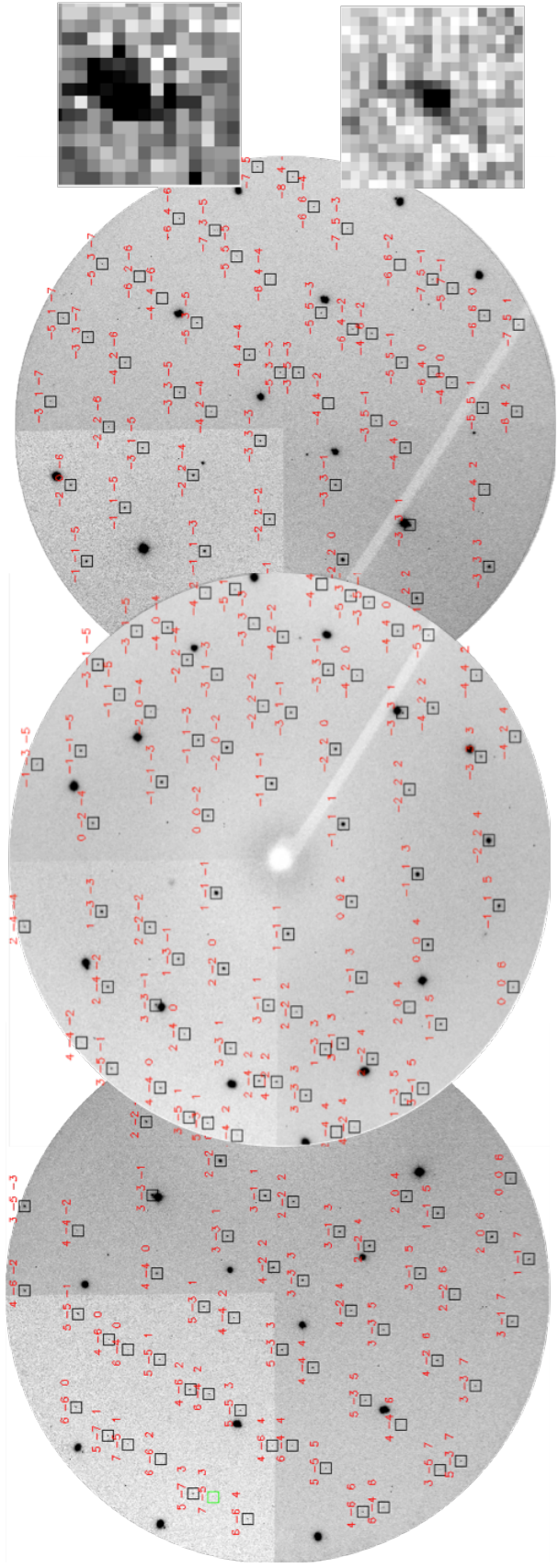

Figure S1. Diffraction images of GaP at ambient pressure in the diamond anvil cell. Peaks are marked with squares and Miller indices. The two zoomed-in peaks on the right side show the highest diffraction angles peaks.

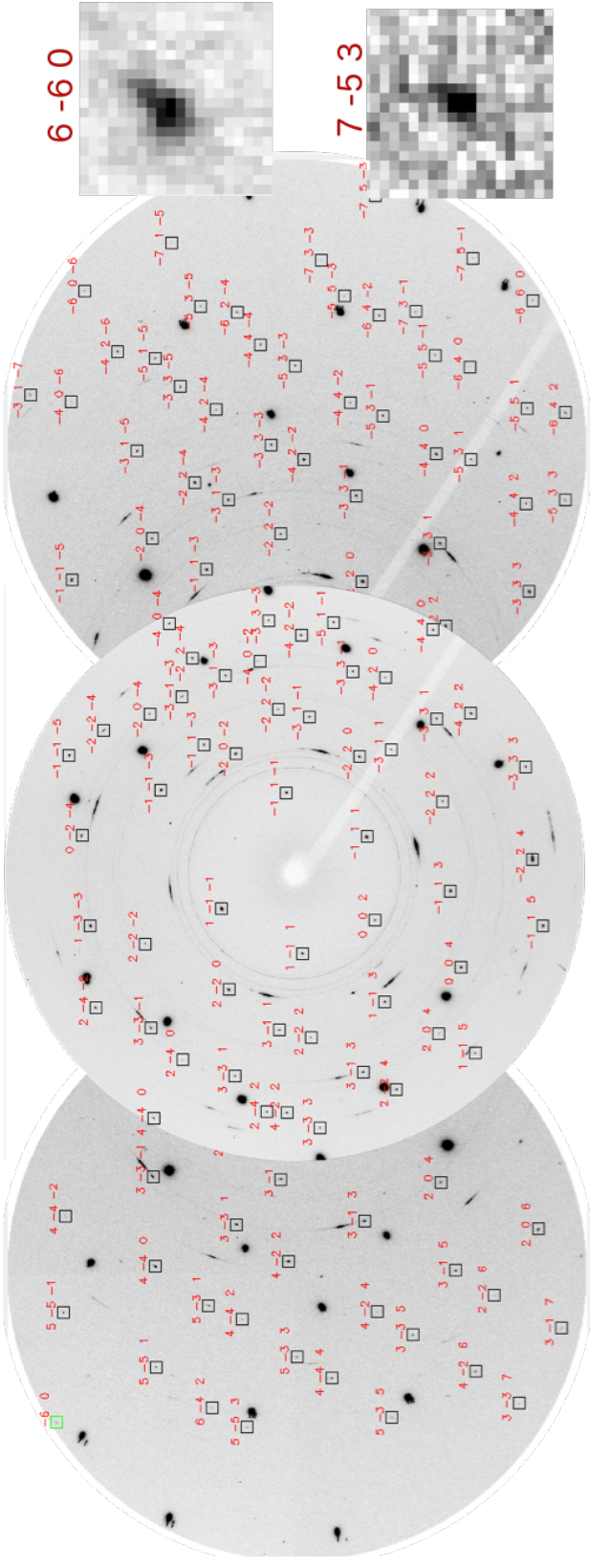

Figure S2. Diffraction pattern of GaP at 24.5 GPa, before the onset of phase transition.

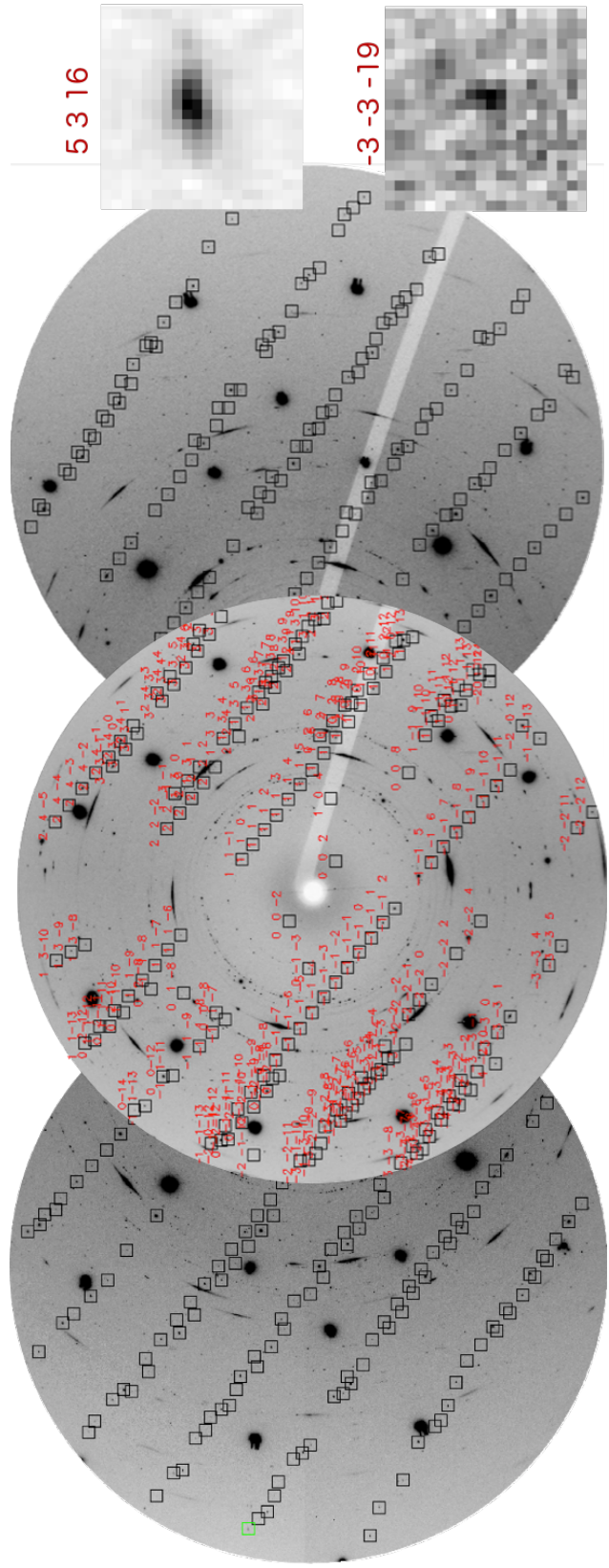

Figure S3. Diffraction pattern of the *oS*24 phase. The peaks of the largest grain are highleghed with squares. Two very high angle peaks are shown to emphasize the high resolution obtained combining laser annealing with multigrain diffraction.

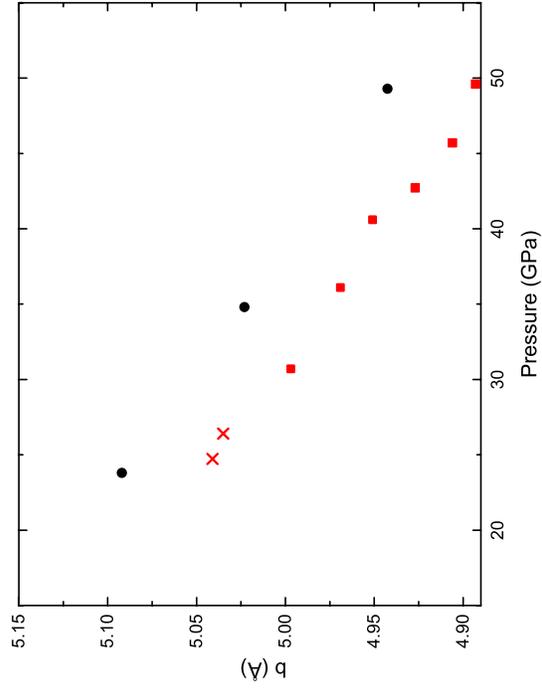
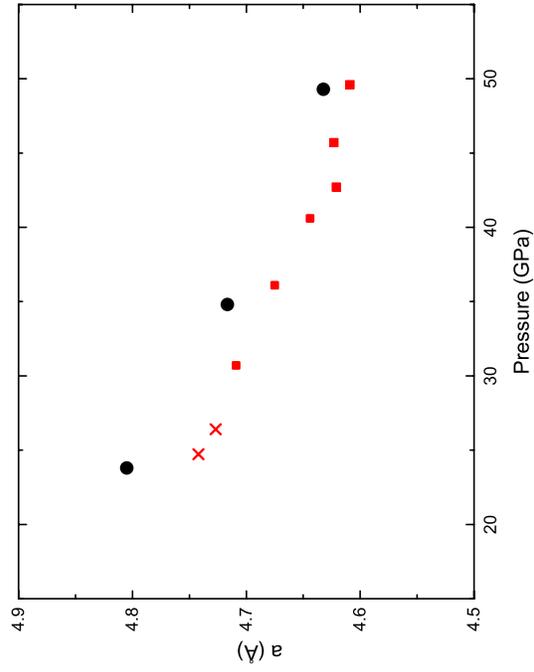
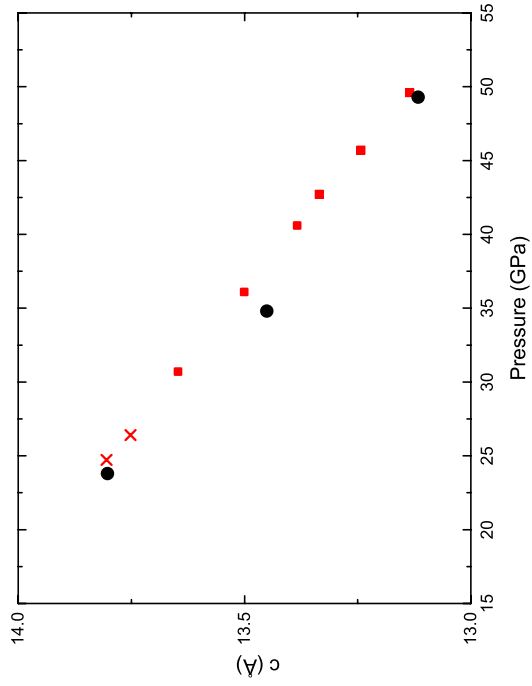

Figure S4. Unit cell parameters of GaP-*oS*24 as a function of pressure. Experimental and calculated values are represented in red and black symbols respectively. Crosses represent datapoints showing weak diffuse scattering lines. Error bars are smaller than symbol sizes.

Table S1 Refined lattice parameters of GaP-*oS*24.

| P(GPa) | a(Å) | b (Å) | c (Å) | Volume (Å³) |
|---|---|---|---|---|
| 42.7 | 4.621(2) | 4.927(2) | 13.335(7) | 303.6(2) |
| 45.7 | 4.623(2) | 4.906(2) | 13.243(2) | 300.4(2) |
| 49.6 | 4.609(3) | 4.893(3) | 13.135(3) | 296.2(3) |
| 40.6 | 4.644(3) | 4.951(3) | 13.384(2) | 307.793 |
| 36.1 | 4.675(3) | 4.969(3) | 13.501(2) | 313.6(3) |
| 30.7 | 4.709(3) | 4.997(3) | 13.647(2) | 321.12(3) |

Table S2 Atomic structure refinement of GaP-*oS*24 at different pressures. Displacement parameters are expressed in Å².

| P(GPa) | $N_{all}$ | N | $R_{eq}$ | $R_\sigma$ | $R_1$ | y(Ga1) | z(Ga1) | $U_{eq}$ (Ga1) | y(Ga2) | $U_{eq}$ (Ga1) | Y(P1) | z(P1) | Ueq (P1) | y(P2) | Ueq (P21) |
|---|---|---|---|---|---|---|---|---|---|---|---|---|---|---|---|
| 42.7 | 324 | 104 | 22 | 13 | 8.7 | 0.1068(5) | 0.0910(3) | [0.010] | 0.4162(7) | [0.010] | 0.6109(13) | 0.0716(6) | [0.006] | 0.104(2) | [0.006] |
| 45.7 | 460 | 153 | 12 | 7 | 4.9 | 0.1072(2) | 0.09076(7) | 0.0083(4) | 0.4138(3) | 0.0087(4) | 0.6110(5) | 0.0718(2) | 0.0058(5) | 0.1081(8) | 0.0065(8) |
| 49.6 | 295 | 97 | 11 | 7.8 | 6.5 | 0.1071(5) | 0.0914(2) | 0.0070(10) | 0.4140(9) | 0.012(1) | 0.6139(12) | 0.0726(5) | [.006] | 0.110(2) | [0.006] |
| 40.6 | 405 | 136 | 14 | 8.5 | 6.3 | 0.1056(3) | 0.09118(7) | 0.0096(5) | 0.4171(4) | 0.0104(4) | 0.6094(6) | 0.0712(2) | 0.0067(6) | 0.1073(9) | 0.0077(10) |
| 36.1 | 454 | 151 | 11 | 7.6 | 6.1 | 0.1036(3) | 0.09140(8) | 0.0116(5) | 0.4198(5) | 0.0116(6) | 0.6071(7) | 0.0709(2) | 0.0076(6) | 0.1042(10) | 0.0092(11) |
| 30.7 | 246 | 79 | 11 | 8.5 | 4.7 | 0.1009(7) | 0.0914(2) | 0.0124(11) | 0.4235(12) | 0.0101(12) | 0.6057(12) | 0.0696(5) | [.006] | 0.100(2) | [.006] |